\documentclass[a4paper,14pt]{extarticle}

\usepackage{graphicx}
\usepackage{amsmath,amsthm,amssymb}

\usepackage[bindingoffset=1cm,
            left=2cm,
            right=2cm,
            top=2cm,
            bottom=2cm,
            footskip=1cm]{geometry}
\usepackage{float}
\usepackage{enumitem}

\title{Sound in a Bubbly Hybrid Neutron Star \thanks{v4 -- 04-04-2024}}
\author{  \textbf{B.O.~Kerbikov} \thanks{bkerbikov@gmail.com} \smallskip \\
 Lebedev Physical Institute,\\ Moscow 119991, Russia \medskip \\
 Moscow Institute of Physics and Technology, \\ Dolgoprudny 141700, Moscow Region, Russia \bigskip \\ \textbf{M.S.~Lukashov} \thanks{m.s.lukashov@gmail.com} \smallskip \\ NRC ``Kurchatov Institute'', \\ Moscow 123182, Russia}

\newcommand{\beq}{\begin{eqnarray}}
 \newcommand{\eeq}{\end{eqnarray}}
\newcommand{\be}{\begin{equation}}
 \newcommand{\ee}{\end{equation}}

\def\fun#1#2{\lower3.6pt\vbox{\baselineskip0pt\lineskip.9pt
\ialign{$\mathsurround=0pt#1\hfil ##\hfil$\crcr#2\crcr\sim\crcr}}}

\newcommand{{\SD}}{\rm SD}

\newcommand{{\Mc}}{\mathcal{M}}

\begin{document}
\maketitle
\begin{abstract}

\noindent Increasingly precise astrophysical observations of the last decade in combination with intense theoretical studies allow for drawing a conclusion about potential Quark Matter presence in Neutron Stars interiors. Quark Matter may form the Neutron Star inner core or be immersed in the form of bubbles, or droplets. We consider the second scenario and demonstrate that even a small fraction of quark matter bubbles can lead to a high nonlinearity of the sound wave. Below the bubble resonant frequency the sound speed is lower than the ambient value. At the resonance it sharply grows. The peak is constrained by viscous dissipation. Above the resonance the speed exceeds the pure neutron star matter value. The dispersion equation for the bubbly neutron star compressibility is derived.





\end{abstract}


\section{Introduction}

In the last years there has been a
breakthrough progress in astrophysical observations on neutron stars (NSs) \cite{01,02,03,04,05,06,07,08}. These advances encourage the studies of the NSs Equation of State (EoS). Construction of the EoS which accounts for the growing amount of astrophysical multi-messenger signals remains largely open problem \cite{09,10,11,12}. The relevant EoS has to include both hadronic matter (HM) and quark matter (QM) degrees of freedom. The presence of QM in NSs is widely discussed and is plausible for maximum masses NSs, see, e.g., \cite{12,13,14,15,16,17,18,19,20,21,22,23,24}. 


In a Hybrid NS (HNS) QM is expected to form the inner core inside the dense HM crust.  This is the most natural but not a unique pattern of the HNS composition. It has been suggested long ago and discussed by a number of authors that the QM insertions of different geometrical structures (drops, rods, slabs, tubes) may be formed inside the HNS \cite{15,16,22,24,25*,26*,25,26,27,28,29,30,31,34*,35*}. The mixed HM-QM state of this type is called pasta phase, see \cite{35*} and references therein. We consider the spherical QM bubbles with an equilibrium radius $R_0$ and leave aside the problem of QM seeds evolution in a process of nucleation or spinodal decomposition \cite{32,33,34}. 

Apart from NSs and HNSs the existence of quark stars (QSs) is presently widely discussed. The basic idea which goes back to \cite{39*,40*,41*} is that QM might be energetically favored over the nuclear matter. Two types of QSs are proposed -- that made of $u$, $d$ and $s$, or $u$ and $d$ quarks, see \cite{42*} and references therein for the first option and \cite{43*} for the second. The core discussion concerns the nature of heavy compact stars with masses $M \leq 2 M_{\odot}$ \cite{42*,44*,45*}. According to very recent analysis of multi-messenger data the maximum mass of NS could be as high as $2.49$ $M_{\odot}$ - $2.52$ $M_{\odot}$ \cite{45*}. The strange QS picture allows to avoid very stiff EoS and consequently to respect the conformal limit of the speed of sound $c_s^2=\dfrac13$ \cite{42*,46*}. On the other hand, according to inference of the multi-messenger observations the NS scenario is favored against the QS scenario \cite{44*}.



An inherent attribute of the EoS is the squared speed of sound $c_s^2=\dfrac{dp}{d\varepsilon}$, where the derivative is considered at constant specific entropy. It describes the stiffness of matter. In \cite{35} it was first clearly indicated that the existence of NSs with masses around two solar masses is in tension with $c_s^2<\dfrac13$ conformal barriere bound. This work gave a start to a flow of publications on non-monotonic behavior of $c_s^2$ as a function of density in NSs, see references above and \cite{36,37,38}. The aim of our work is to investigate the sound propagation in HNS with QM droplets (bubbles) immersed into it. It will be shown that the presence of QM bubbles in HM causes a highly nonlinear behaviour of the sound wave propagation. The dispersion curves of the sound phase speed and of the rate of attenuation exhibit a dip-bump structure in the neighbourhood of the QM bubble resonance frequency. Above the resonance the stiffness of the matter increases and the sound speed becomes higher than in the pure HM. The anomalous dispersion of the sound phase speed in a bubbly liquid has been studied by several authors, see, e.g., \cite{39,40}, and will be investigated below using the Rayleigh-Plesset (RPE) equation $\cite{41,42}$ as a starting point. To describe the sound propagation in a bubbly HNS one needs the EoS's of HM and QM. To this end we shall rely on the model-independent polytrope EoS proposed in \cite{18}. It has proved itself in successfully fitting the astrophysical data on NSs \cite{07,12,43}.

The work is organized as follows. In Sec. 2 we remind the Rayleigh equation (RE) and present its generalizations with the inclusion of the dissipation, driving pressure, surface tension and polytropic pressure-volume relation. In Sec. 3 the equation for the bubble volume response to the oscillatory driving pressure is derived. The relationship between relativistic and the Newtonian (adiabatic) polytropes is established. Sec. 4 is the core of the paper. The formula for the compressibility of the HM containing QM bubbles is derived. Expressions for the sound speed and attenuation coefficient in a bubbly HNS are presented. The choice of parameters characterising the HNS is discussed in Sec. 5. In Sec. 6 we set up the firm ties between different parameters. The results of the calculations of the speed of sound and the attenuation coefficient are presented in Sec. 7. Sec. 8 contains the summary of the work. Throughout the work, we follow the condition $\hbar = c = 1$.

\section{Rayleigh and Rayleigh-Plesset Equations}

The theory of a bubble dynamics in an infinite body of liquid dates back to the work of Lord Rayleigh \cite{41}. In 1917 he investigated cavitation damage of the ship propellers and discovered that it was caused by the bubbles collapse. The RE describing the bubble pulsation and collapse (inertia cavitation) reads 

\be R \ddot{R}+\frac{3}{2} \dot{R}^2=0, \label{01}\ee 

where $R(t)$ is the bubble radius. The solution of (\ref{01}) is the power law 

\be R(t) \sim\left(t_c-t\right)^{2 / 5}, \label{02} \ee 

which leads to a divergent wall velocity $\dot{R}(t) \sim \left(t_c-t\right)^{-3 / 5}$ at $t \rightarrow t_c$. The RE (\ref{01}) gives an oversimplified picture of bubble phenomena with only inertia forces accounted for. The simplest generalization of RE is the celebrated RPE \cite{42,44,45} which reads

\be
R \ddot{R}+\frac{3}{2} \dot{R}^2=\frac{1}{\rho_h}\left\{p_q-p_h-\frac{2 \sigma}{R}-4 \eta \frac{\dot{R}}{R}-P(t)\right\}.
\label{03}\ee

\noindent where $\rho_h$ is the density of the medium surrounding the bubble (the HM density), $p_q$ is steady and unsteady pressures in the QM bubble interior, $p_h$ is the undisturbed ambient HM pressure, $\sigma$ is the surface tension, $\eta$ is the surrounding HM shear viscosity, $P(t)$ is the driving acoustic pressure. A pedagogical derivation of (\ref{03}) may be found in \cite{01*,46}. There are only a few studies of QM bubble dynamics based on (\ref{01}) and (\ref{03}) \cite{46,47,48}. In application to bubbly HNS both $\sigma$ and $\eta$ play an important role. It will be shown that the oscillation regime around the equilibrium radius is realized provided $\sigma$ is below a certain critical value. We note in passing that the critical value of $\sigma$ has been discussed in literature \cite{24,26,27,35*} in relation to the character of the QM-HM transition. The shear viscosity possibly prevents the collapse of the QM bubble \cite{47} and puts the casual upper limit on the speed of sound, see below. From what follows it will be clear that the values of both parameters are poorly known, not to say at a level of an educated guess.


One can notice that (\ref{03}) contains the HM shear viscosity $\eta$ but bulk viscosity $\zeta$ is absent. Without going into details we indicate the assumption which led to the elimination of $\zeta$. The RPE is based on the viscous Navier-Stokes equation (NSE) and the boundary condition on the bubble wall \cite{01*,46,83}. The NSE contains both $\eta$ and $\zeta$ \cite{01*,46}. The boundary condition is the matching equation for the radial component of the stress tensor $\sigma_{rr}$ at the bubble interface \cite{01*,46,83}. The stress tensor $\sigma_{rr}$ reads \cite{02*}

\be
\sigma_{rr}=-p_L+2 \eta \frac{\partial v}{\partial r}+\left(\zeta-\frac{2 \eta}{3}\right) \mathbf{\nabla} \mathbf{v} \label{04*}
\ee

where $p_L$ is the HM pressure at the outer interface of the bubble and $ v(r,t)=\frac{R^2(t)}{r^2(t)} \dot{R}(t)$ \cite{01*,46}.

To derive RPE in the form (\ref{03}) one has to assume that the motion of HM at the bubble wall is incompressible, that is, $\mathbf{\nabla} \mathbf{v}=\operatorname{div} \mathbf{v}=0$ in (\ref{04*}). Outside the wall HM compressibility is responsible for the undisturbed speed of sound $c_h$. After this assumption is made the derivation of RPE proceeds as described in \cite{46}. The compressibility corrections to RPE were considered in \cite{49}. It was shown that RPE has the error of the order of the bubble wall Mach number $c_h^{-1} d R / d t$, where $c_h$ is the speed of sound in the surrounding HM. The Mach number grows in the vicinity of the bubble collapse. One may expect that bulk viscosity will play an important role in this region as well. It is known that $\zeta$ has a maximum close to the second-order phase transition temperature and near the QCD critical endpoint \cite{03*,04*,05*}. 

It is also enhanced in presence of the slow relaxation Mandelstam \!-\! Leontovich mode \cite{04*}. Calculations have shown, see \cite{71} and references therein that bulk viscosity of NSs depends on density, temperature, mechanisms accounted for, etc. Very roughly speaking, the shear viscosity $\eta$ dominates over the bulk viscosity $\zeta$ \cite{72*}.

Strictly speaking, RPE (\ref{03}) is correct for small deviations of the ball radius and pressure from their equilibrium values \cite{49}. This opens the possibility to describe the acoustic properties of bubbly HNSs using RPE and the polytropic EoS.

\section{Oscillations of a QM Bubble with a Polytropic EOS}

Consider (\ref{03}) with small deviations of the bubble radius and pressure from their equilibrium values $R_0$ and $p_{q0}$: $R=R_0[1+x(t)]$, $p_q=p_{q0}+r$, where $r$ is the variable part of the pressure inside the bubble (not to be confused with the range $r$ from the bubble center in (\ref{04*})). We insert the above expressions for $R$ and $p_q$ into (\ref{03}) and perform linearization. One gets

\be
\ddot{x}=\frac{1}{\rho_h R_0^2}\left\{r+\frac{2 \sigma}{R_0} x-4 \eta \dot{x}-P(t)\right\} .
\label{04}
\ee

Linearization means neglecting terms proportional to $x \ddot{x}$, $\dot{x}^2$, $x^2$. This is legitimate since we consider bubble oscillations with small amplitude $x \sim a e^{i \omega t}$, $a \ll 1$. Then $R \ddot{R} \simeq R_0^2 \ddot{x}+R_0^2 {x} \ddot{x}$, $\ddot{x} \sim a$, $\quad x \ddot{x} \sim a^2$, $\quad x \ddot{x} / \ddot{x} \sim a<1$. Similar argument applies to other terms. An important point is that small bubble oscillations result in highly nonlinear sound propagation as we shall see below.

Next comes the polytropic EoS \cite{18,07,43,50} which in a model-independent way describes HM and QM components of HNS and allows to relate $r$ to $x$. The EoS \cite{18} and a family of piecewise EoSs generated from it meet the multimessenger picture of NSs \cite{07}.

According to \cite{18} the EOS is formulated in terms of the polytropic index defined as
\be 
\gamma=d(\ln p) / d(\ln \varepsilon) = \frac{\varepsilon}{p} \frac{d p}{d \varepsilon}=\frac{\varepsilon}{p} c_s^2, 
\label{ex37}\ee 

where $\varepsilon$ is the energy density. The index $\gamma$ takes the value $\gamma \approx 2.5$ around the saturation density, $\gamma=1.75$ is the HM-QM deviding line, $\gamma \rightarrow 1$ in high density QM, $\gamma<0.5$ destabilizes the star \cite{18,07,55,70}. In RPE one has to resort to the polytropic EoS expressing pressure as a function of density, $p=k \rho^{\Gamma}$ with ${\Gamma}$ usually called the adiabatic index. In what follows we change the notation from $\Gamma$ to $\bar{\gamma}$ which is more handy to use in formulas. Interconnection of the two polytrophic forms is discussed at the end of this Section. In terms of the bubble radius $R$ the last polytrope reads

\be
p_q R^{3 \bar{\gamma}}=p_{q0} R_0^{3 \bar{\gamma}}.
\label{05}
\ee

Next we expand $(1+x)^{3 \bar{\gamma}} \approx 1 + 3 \bar{\gamma} x$ and obtain $r = -3 \bar{\gamma} p_{q0} x$. Then (\ref{04}) takes the form

\be
\ddot{x}=\frac{1}{\rho_h R_0^2}\left\{-\left(3 \bar{\gamma}_q p_{q0}-\frac{2 \sigma}{R_0}\right) x - 4 \eta \dot{x}-P(t) \right\} .
\label{06}\ee

We assume the oscillatory driving excitation pressure $P(t)=p_s e^{i \omega t}$. Then (\ref{06}) may be rewritten as

\be
\ddot{x}+g \dot{x}+\omega_0^2 x=-\frac{p_s}{\rho_h R_0^2} e^{i \omega t}, 
\label{07} \ee

\be
g=\frac{4 \eta}{\rho_h R_0^2},\qquad \omega_0^2=\frac{1}{\rho_h R_0^2}\left(3 \bar{\gamma}_q p_{q0}-\frac{2 \sigma}{R_0}\right).
\label{08} \ee

We see that (\ref{07}) is an equation of a damped forced harmonic oscillator with frequency $\omega_0^2$ and viscous friction damping $g$. It admits an analytical solution to be presented below. The requirement of positive stiffness $\omega_0^2$ imposes the upper bound on $\sigma$ for given $R_0$ and other parameters. The stability condition reads

\be
\varphi = \frac{2 \sigma}{R_0} ( 3 \bar{\gamma}_q p_{q0} )^{-1} < 1.
\label{09} \ee

We shall return to the discussion of this relation later in Section 5. For future purposes we rewrite (\ref{07}) in the volume frame. In linear approximation the dynamical volume $v$ is

\be
v \equiv V-V_0 \simeq \frac{4}{3} \pi R_0^3 (1 + 3 x)-\frac{4}{3} \pi R_0^3 = 3 V_0 x.
\label{10} \ee

We note that it does not make sense to go beyond the approximation (\ref{10}) with poorly known values of physical parameters. In terms of $v$ equation (\ref{07}) takes the form

\be
\ddot{v}+g \dot{v}+\omega_0^2 v=-d p_s e^{i \omega t},
\label{11} \ee

with $d=\dfrac{4 \pi R_0}{\rho_h}$. We note that relativistic generalization of RPE (\ref{03}) and hence of (\ref{11}) is straightforward \cite{48,51} but the sound propagation equations become less transparent.

Now we return to the relationship between the two polytropic indices $\gamma$ and $\bar{\gamma}$. One can recast the polytrope $\bar{\gamma}$ to meet with (\ref{ex37})

\be \bar{\gamma}=d(\ln p) / d(\ln \rho)=\frac{\rho}{p} \frac{d p}{d \rho}=\frac{\rho}{p} \bar{c}_s^2.\label{ex39}\ee

The polytropes (\ref{ex37}) and (\ref{ex39}) are called relativistic and Newtonian \cite{65} correspondingly. Note that in a classical textbook \cite{52} $\bar{c}_s$ defined in (\ref{ex39}) is called the speed of sound. The relation between (\ref{ex37}) and (\ref{ex39}) has been discussed by a number of authors \cite{65,61,62,63,64,66}. Assuming that the energy density $\varepsilon$ and the pressure $p$ are functions of density $\rho$ only, one can write the first law of thermodynamic as $dE+pdV=0$. Together with $\varepsilon=E / V$, $\rho = M / V$, and $d (1 / \rho) = dV / M$ it leads to \cite{65,64}

\be
d\left(\frac{\varepsilon}{\rho}\right)=-p d\left(\frac{1}{\rho}\right) .
\label{38*}
\ee

Integration of (\ref{38*}) with respect to $\rho$ with the account of the nonrelativistic limit $\varepsilon = \rho$ yields the desired equation

\be
\varepsilon=\rho+\frac{1}{\bar{\gamma}-1} p.
\label{39*}
\ee

From (\ref{ex37}), (\ref{ex39}) and (\ref{39*}) one easily gets

\be
\bar{\gamma}=\frac{\varepsilon + p}{\varepsilon} \gamma.
\label{40*}
\ee

According to \cite{61} the linear dependence of the type (\ref{39*}) connecting $\varepsilon$ to $\rho$ is valid for a wide class of liquids. However, the above relations are purely thermodynamic. As an example of $\varepsilon / \rho$ connection based on dynamical models we refer to Fig. 5 of \cite{67}.

To summarize, it does not make much difference whether $\gamma$ or $\bar{\gamma}$ is implied in the set of parameters presented below. In what follows we shall keep notation $\bar{\gamma}$ for the polytrope (\ref{ex39}) but omit the vertical bar from $\bar{c}_s$.

\section{Sound in a Bubbly Hybrid Neutron Star}

Assuming spatially uniform distribution of $N$ bubbles with equal volume $V$ we write the average density $\rho_m$ of QM-HM mixture as

\be
\rho_m=(1-\beta) \rho_h+\beta \rho_q.
\label{12} \ee

Here $\rho_h$ and $\rho_q$ are the densities of HM and QM components, $\beta$ is the bubble volume fraction $\beta = NV/\left(U_h+NV\right)$, $U_h$ is the HM volume, the total volume is $U=U_h+NV$. The QM volume fraction $\beta$ is assumed to be small, $\beta \ll 1$. To find the speed of the pressure wave in the mixture we differentiate (\ref{12}) with respect to the insonifying field pressure. This task turns out to be far from trivial. The bubble volume $V=V_0+v$ "breathes" with small amplitudes according to (\ref{11}). The bubble volume fraction $\beta$ becomes pressure-dependent as well. Differentiation of (\ref{12}) yields

\be
\frac{d \rho_m}{d p}=\frac{1}{c_m^2}=\frac{1-\beta}{c_h^2}+\beta \frac{d \rho_q}{d p}-\rho_h \frac{d \beta}{d p}+\rho_q \frac{d \beta}{d p}
\label{13} \ee

Here the sound velocity is $c_s^2 = d p / d \rho$, where the derivative is considered at constant specific entropy. Comparison with $c_s^2 = d p / d \varepsilon$ is discussed in Sec. V. With $U_q=NV$ being the total QM volume, $M_q$ its mass, we have

\be
\frac{d \rho_q}{d p}=\frac{d}{d p} \frac{M_q}{U_q}=-\left(\frac{M_q}{U_q}\right)\left(\frac{N}{U_q}\right) \frac{d v}{d p}=\rho_q\left(-\frac{1}{V} \frac{d v}{d p}\right)=\rho_q \varkappa_q. \label{14} \ee

Here $\varkappa_q = -\frac{1}{V} \frac{d v}{d p}$ is the compressibility of QM. In a similar way $d \beta / d p$ is evaluated

$$
\frac{d \beta}{d p}=\frac{d}{d p} \frac{U_q}{U_h+U_q}=\frac{N}{U_h+U_q} \frac{d v}{d p}-\frac{U_q N}{\left(U_h+U_q\right)^2} \frac{d v}{d p}=
$$

\be
=-\beta \varkappa_q+\beta^2 \varkappa_q.
\label{15}
\ee

The term proportional to $\beta^2$ can be dropped, the second and the last terms in (\ref{13}) cancel each other and we arrive at
\be
\frac{1}{c_m^2} \cong \frac{1-\beta}{c_h^2}+\beta \rho_h \varkappa_q . \label{16}
\ee
In general, the speed of sound is expressed in terms of  compressibility as follows

\be
\varkappa=-\frac{1}{V} \frac{d V}{d p}=-\rho \frac{d(1 / \rho)}{d p}=\frac{1}{\rho} \frac{d \rho}{d p}=\frac{1}{\rho c_s^2},(17)
\label{17}
\ee

so that $c_s^2=1 / \rho \varkappa$. To find $\varkappa_q$ in (\ref{16}) one has to solve (\ref{11}) for $v$. Taking $v$ in the form $v=v^{\prime} e^{i \omega t}$ this is easily done with the result

\be
v^{\prime}=-\frac{3 V_0 p_s}{\rho_h R_0^2\left(\omega_0^2-\omega^2+i g \omega\right)}.
\label{18}
\ee

The sound wave pressure has the oscillatory form $p=p_s e^{i \omega t}$. Therefore

\be
\varkappa_q=-\frac{1}{V_0} \frac{d v}{d p}=-\frac{1}{V_0} \frac{d v}{d t}\left(\frac{d p}{d t}\right)^{-1}=-\frac{1}{V_0} \frac{v^{\prime}}{p_s}.
\label{19}
\ee

Insertion of (\ref{18}) into (\ref{19}) yields

\be
\varkappa_q=\frac{3}{\rho_h R_0^2}\left(\omega_0^2-\omega^2+i g \omega\right)^{-1}.
\label{20}
\ee

Returning to (\ref{16}) we write

\be
\frac{c_h^2}{c_m^2}=1+\frac{3 \beta c_h^2}{R_0^2}\left(\omega_0^2-\omega^2+i g \omega\right)^{-1}.
\label{21}
\ee

A comment on terminology is appropriate at this point. Equations (\ref{03}), (\ref{06}), (\ref{11}) contain the shear viscosity $\eta$. The QM bubble oscillates with friction which leads to the sound attenuation. As a result, $\varkappa_q$ and $c_m$ are complex while according to the classical textbook \cite{52} the sound velocity is real by its meaning. This contradiction a fictitious. The sound phase speed to be obtained below is real. In a system with dissipation the sound wavenumber is complex. However, it is quite common to call speed of sound the complex quantities like $c_m$ \cite{53,54,55}.

The relation (\ref{18}) enables to obtain the real phase speed and the attenuation coefficient. The pressure wave propagates in a bubbly medium with a complex wavenumber $k_m=k_1+i k_2$ and has the form 

\be
u=A e^{i\left(\omega t - k_m x\right)}=A e^{i \omega\left(t-\frac{k_1}{\omega} x\right)} e^{k_2 x}=A e^{i \omega\left(t-\frac{x}{c_{ph}}\right)} e^{-\alpha x},
\label{22}
\ee

where $c_{ph}$ is the phase speed and $\alpha$ is the absorption coefficient

\be
c_{ph}=\frac{\omega}{\operatorname{Re} k_m},\quad \alpha=- \operatorname{Im} k_m .
\label{23}
\ee

To express $c_{ph}$ and $\alpha$ in terms of HM-QM parameters we set

\be
\frac{c_h}{c_m}=\frac{k_m}{k_h}=\nu - i \xi, \quad \frac{c_h^2}{c_m^2}= G - i F.
\label{24}
\ee

Simple algebra leads to

\be
c_{ph}=\frac{c_h}{\nu}, \quad \alpha=\frac{\omega}{c_h} \xi = \omega \frac{F}{2 c_h \nu},
\label{25}
\ee

\be
\nu^2=\frac{1}{2}\left(G+\sqrt{G^2+F^2}\right) .
\label{26}
\ee

We remind that $c_h^2=\frac{\gamma_h p_h}{\rho_h}$, $\gamma_h > 1.75$.

The ratio (\ref{24}) was already evaluated and is given by (\ref{21}). From (\ref{21}) we get


\be
G=1+\frac{3 \beta c_h^2}{R_0^2} \frac{\omega_0^2-\omega^2}{\left(\omega_0^2-\omega^2\right)^2+g^2 \omega^2} \\
\label{27}
\ee

\be
F=\frac{3 \beta c_h^2}{R_0^2} \frac{g \omega}{\left(\omega_0^2-\omega^2\right)^2+g^2 \omega^2} .
\label{28}
\ee

Equations (\ref{25})-(\ref{28}) provide a complete solution for the sound phase speed and the attenuation coefficient. 

At this point the values of the physical prosameters entering into (\ref{27})-(\ref{28}) remain unfixed except for the requirement $\beta \ll 1$. The choice of parameters is discussed below in Section 5. It is instructive to separate in $\omega_0^2$ the surface tension contribution. We remind equation (\ref{08}) for $\omega_0^2$ and present it in the following form

\be
\omega_0^2=\bar{\omega}^2_0 \Phi^2, \quad \bar{\omega}^2_0 = \dfrac{3 \bar{\gamma}_q p_{q 0}}{\rho_h R^2_0}, \quad \Phi^2=1-\varphi, \quad \varphi=\dfrac{2 \sigma}{3 \bar{\gamma}_q p_{q o} R_0}.
\label{30} \ee

Then (\ref{27})-(\ref{28}) read

\be G=1+\beta \frac{\bar{\gamma}_h p_h}{\bar{\gamma}_q p_{q o}} \frac{\Phi^2-\Omega^2}{\left[\left(\Phi^2-\Omega^2\right)^2 + \delta^2 \Omega^2\right]},
\label{31} \ee

\be F=\beta \omega \frac{\bar{\gamma}_h p_h}{\bar{\gamma}_q p_{q o}} \frac{g / \bar{\omega}_0^2}{\left[\left(\Phi^2-\Omega^2\right)^2 + \delta^2 \Omega^2\right]},
\label{32} \ee

\noindent where $\Omega=\omega / \bar{\omega}_0$, $\delta=g / \bar{\omega}_0$. From (\ref{25})-(\ref{26}) and (\ref{31})-(\ref{32}) one easily obtains the sound velocity and the attenuation rate

\be c_{\rho h} \simeq \frac{c_h}{\sqrt{G}}=c_h\left(1+\beta \,\frac{\bar{\gamma}_h p_h}{\bar{\gamma}_q p_{q0}} \,\frac{\Phi^2-\Omega^2}{\left[\left(\Phi^2-\Omega^2\right)^2+\delta^2 \Omega^2\right]}\right)^{-\frac12},
\label{33} \ee

\be \alpha=\frac{\omega}{2 c_h F}=\beta\left(\frac{4}{3}\,\eta\,\frac{\omega^2}{2 \rho_q {c}_q^3}\right) \frac{\rho_h c_h}{\rho_q {c}_q} \, \frac{1}{\left[ \left(\Phi^2-\Omega^2\right)^2+\delta^2 \Omega^2 \right]},
\label{34} \ee

\noindent where ${c}_q=\sqrt{\frac{\bar{\gamma}_q p_{q0}}{\rho_q}}$. Interesting to note that $\alpha$ contains a factor

\be
\alpha^{\prime}=\frac{4}{3}\,\eta\,\frac{\omega^2}{2 \rho_q {c}_q^3},
\label{35} \ee

\noindent which is a well known sound attenuation coefficient at zero bulk viscosity and zero thermal conductivity \cite{52}. The absence of the bulk viscosity $\zeta$ in (\ref{35}) was explained at the end of Section 2. In the low frequency limit $\omega \rightarrow 0$ (\ref{33}) and (\ref{34}) reduce to

\be c_{p h} \simeq c_h\left(1+\frac{\beta}{\Phi}\,\frac{c_h^2 \rho_h}{{c}_q^2 \rho_q}\right)^{-\frac12} \simeq c_h\left(1-\frac{\beta}{\Phi}\,\frac{c_h^2 \rho_h}{{c}_q^2 \rho_q}\right),
\label{36}\ee
\be \alpha \simeq \alpha^{\prime} \frac{\beta}{\Phi^2}\,\frac{\rho_h c_h}{\rho_q {c}_q}.
\label{37}\ee

The dependence of (\ref{33})-(\ref{34}) and (\ref{36})-(\ref{37}) on $\Phi^2=1-\varphi$ exhibits two limiting regimes. The first one is $\varphi \rightarrow 0$, $\Phi^2 \rightarrow 1$ which is realized for small values of $\sigma$ and/or for large bubble radius. In this care one simply replaces $\Phi^2$ by 1 in the above equations. The opposite limit $\Phi^2 \rightarrow 0$, $\varphi \rightarrow 1$ corresponds to vanishing resonance frequency $\omega_0$, where $\omega_0^2=\bar{\omega}_0^2(1-\varphi)$. The neighbourhood of the bubble stability limit $\varphi \rightarrow 1$, $\omega_0^2 \rightarrow 0$ deserves a special consideration. At $\omega_0^2 \to 0$ the quantities $G$ (\ref{31}) and $F$ (\ref{32}) read

\be G=1-\frac{3 \beta c_h^2}{R_0^2}\,\frac{1}{\omega^2+g^2},
\label{38} \ee

\be F=\frac{3 \beta c_h^2}{R_0^2}\,\frac{g / \omega}{\omega^2+g^2}.
\label{39} \ee

To get the sound speed and the attenuation coefficient one has to return to (\ref{25})-(\ref{26}). The approximation $\nu \simeq \sqrt{G}$ valid for $F^2 / G^2 \ll 1$ leading to (\ref{33})-(\ref{34}) is broken for $\varphi \rightarrow 1$, $\omega_0^2 \rightarrow 0$ due to the factor $\omega^{-1}$ in $F$. The most interesting case is when $G$ is negative. Then $\nu^2 \rightarrow F^2 / 4|G|$ and

\be c_{ph} \simeq \frac{2 \sqrt{|G|}}{F} c_h=\frac{2 \omega g R_0}{\sqrt{3 \beta\left(\omega^2+g^2\right)}},
\label{40} \ee

\be \alpha \simeq \frac{\omega \sqrt{|G|}}{c_h}=\frac{\omega}{R_0} \sqrt{\frac{3 \beta}{\omega^2+g^2}}.
\label{41} \ee

From (\ref{33})-(\ref{34}) and (\ref{40})-(\ref{41}) we conclude that the sound velocity and attenuation in a bubbly HNS depend on the surface tension at the QM-HM interface and on the shear viscosity of the HM.

\section{The Choice of Parameters}

The values of physical parameters entering into (\ref{33})-(\ref{34}) are highly uncertain. We take for granted the polytropic EoS $\gamma=d(\ln p) / d(\ln \varepsilon)$ \cite{18} with $\gamma=1.75$ the dividing line between HM and QM. As explained in Section 3 in RPE one uses the polytrope $\rho=k \rho^{\bar{\gamma}}$ where $\bar{\gamma}$ is the so-called adiabatic index. The linear relation between $\gamma$ and $\bar{\gamma}$ is given by (\ref{41}). We identify $\bar{\gamma}_c=1.75$ as a value dividing the two phases, $\bar{\gamma}_q < 1.75$ corresponds to the QM phase and $\bar{\gamma}_h > 1.75$ to the HM one.

Irrespective of attribution of $\bar{\gamma}$ values to different HNS phases, the range of other parameters is wide, model dependent and loosely defined by observations. In addition, one should specify the NS under consideration and the distance from the QM bubble to the NS center. Density and pressure strongly depend upon this distance. Therefore we take some tentative values of the parameters lying within the interval accepted in most papers on the subject. Our aim is to display the qualitative picture of the sound dispersion and attenuation in a bubbly HNS.

Due to continuous progress in NSs observations and intense theoretical work the possible set of parameters presented in the literature is rather diverse, see, e.g., the recent review \cite{a}. Apart from the $\bar{\gamma}$ values the other three key parameters are the sound velocity, density and pressure. We take for them the values within the bands of the number of solutions without sticking to a particular one. Our set of parameters is the following. For HM the values are 

\be \bar{\gamma}_h=2.5, \quad c_{sh}^2=\frac{1}{3}, \quad \rho_h=250 \frac{\mathrm{MeV}}{\mathrm{fm}^3}=\frac{5}{3} \rho_0, \quad p_h=33 \frac{\mathrm{MeV}}{\mathrm{fm}^3}.
\label{42} \ee

For QM we choose

\be \bar{\gamma}_q=1.4, \quad c_{sq}^2=\frac{1}{2}, \quad \rho_q=600 \frac{\mathrm{MeV}}{\mathrm{fm}^3}=4 \rho_0, \quad p_{q0}=214 \frac{\mathrm{MeV}}{\mathrm{fm}^3}.
\label{43} \ee

Here $\rho_0=150$ $\frac{\mathrm{MeV}}{\mathrm{fm}^3}$ is the saturation density corresponding to $n_0=0.16$ fm$^{-3}$. The above numbers were choosen in the following way. First, in line with \cite{18} we fix the values of $\bar{\gamma}_h$ and $\bar{\gamma}_q$ on different sides of $\bar{\gamma}_c=1.75$. Then comes the choice of the sound speed. For HM we take the conformal limit value $c_{sh}^2=1 / 3$. From a number of publications it is known that $c_{sh}^2$ exhibits bumps and wiggles (see, e.g., \cite{a,b}) but a monotonous solution around the conformal value is not excluded \cite{50}. One can take $c_{sh}^2=1 / 3$ as a guide. As it was first pointed out in \cite{35} and confirmed by many authors the conformal limit is surpassed in QM unless the density is asymptotically high. We choose for $c_{sq}^2$ the value $c_{sq}^2=1 / 2$. With the values of $\bar{\gamma}_c$, $\bar{\gamma}_h$, $\bar{\gamma}_q$, $c_{sh}^2$, $c_{sq}^2$ at hand we sample the density and the pressure values. The density at which QM possibly appears in HNS is very uncertain and depends on the star mass. According to \cite{08} it may occur at (2-3)$n_0$, in \cite{b} the importance of quark degrees of freedom is expected at (2-4)$n_0$, while \cite{09} attributes the HM-QM transition to densities (5-7)$n_0$.

The density values in (\ref{42})-(\ref{43}) meet the (2-4)$n_0$ criterion. The corresponding pressure values cannot be uniquely fixed. As an example we refer to Fig. 1 of \cite{a}. Our pressure values are within the intervals depicted in this figure. Three more parameters are needed to calculate the sound velocity (\ref{33}) and damping (\ref{34}). They are the bubble radius $R_0$, the surface tension $\sigma$ and the HM viscosity $\eta$. The values of these quantities are almost unrestricted. The bubble radius may be to some extent considered as a free parameter though according to a recent study the preferable value is $R_0 \sim(5-10)$ fm \cite{35*}. Nucleation of QM drops with much larger $R$ has been discussed in \cite{73,75}. The surface tension $\sigma$ is a parameter closely connected with the question of whether QM-HM phase transition is a sharp first-order or a smooth one \cite{24,26,27,28,35*,56,57,58}. The value of $\sigma$ was recently discussed in \cite{35*} and according to this reference the value of $\sigma$ spans from $10$ {MeV}/{fm}$^2$ to $120$ {MeV}/{fm}$^2$. The determination of $\sigma$ is a complicated model-dependent task which is beyond our subject. Shear viscosity $\eta$ leads to the sound attenuation and prevents the sound speed to become acausal. Shear viscosity plays an important role in heavy-ion collisions. However, a direct measurements of $\eta$ are not possible and observables are expressed in terms of the ratio of $\eta$ over the entropy density, $\eta / s$. A theoretical lower bound (KSS bound) is $\eta / s \geqslant \frac{1}{4 \pi}$ \cite{68}. Calculations of $\eta$ in NSs \cite{71,69,70,72} take into account the contributions from leptons, neutrons and effects from superfluidity. The results strongly depend on temperature and density and are presented in terms of $\log _{10} \eta \left[ \text{g\,cm}^{-1}\text{s}^{-1} \right]$. The results are within a wide range from $10$ to $20$. All three parameters $R_0$, $\sigma$ and $\eta$ are interrelated. This is discussed in the next Section.

\section{Bringing Parameters Together}

It makes sense to present a list of relations linking the key parameters which determine the speed of sound and attenuation. Some of these equations were already pretended above.

\be \dfrac{2 \sigma}{R_0}=p_{q 0}-p_h,
\label{44} \ee

\be \varphi=\dfrac{2 \sigma}{R_0} \dfrac{1}{3 \bar{\gamma}_q p_{q 0}},
\label{45} \ee

\be \delta=g / \bar{\omega}_0=\dfrac{4}{\sqrt{3 \bar{\gamma}_q \rho_h \rho_{q 0}}}\left( \dfrac{\eta}{R_0} \right).
\label{46} \ee

Here (\ref{44}) is the Young-Laplace pressure equation, $R_0$ is the bubble radius in equilibrium, $p_{q0}$ is the static inside pressure at $R=R_0$ in absence of any driving perturbations. The set (\ref{44})-(\ref{46}) may be supplemented by the expression for the resonance sound wave frequency

\be f_r=\dfrac{1}{2\pi} \bar{\omega}_0 \sqrt{1 - \varphi}=\dfrac{1}{2 \pi R_0} \sqrt{\dfrac{1}{\rho_h}\left(3 \bar{\gamma}_q p_{q_0}-\dfrac{2 \sigma}{R_0}\right)}.
\label{47} \ee

This equation is important in view of a rapid development of the NSs seismology. The canonical NSs frequencies are in the kHz range \cite{76,77} but recently the MHz range signals were discussed \cite{78}. There might be different sources of sound in NSs like, e.g., phase transition \cite{999,04*}.

\begin{figure}[h]
\begin{center}
\includegraphics[width=140mm,keepaspectratio=true]{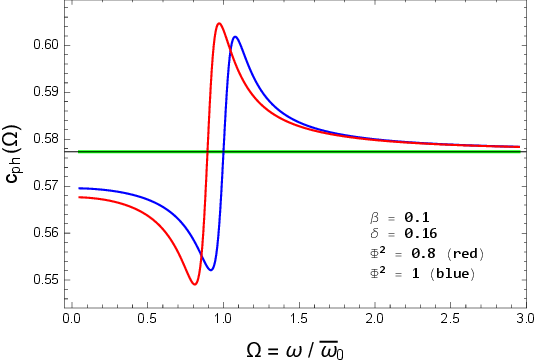}
\caption{The speed of sound in bubbly HNS as a function of $\Omega$ at $\beta=0.1$, $\delta = 0.16$. The red one corresponds for $\Phi^2=0.8$, and the blue one -- for $\Phi^2=1$. The green line is the sound speed with no bubbles.}
\end{center}
\label{figure01}
\end{figure}

As explained above, the values of the parameters in (\ref{44})-(\ref{47}) are known very loosely and may be shifted remaining within the range of the multimessenger solutions. Therefore the solution presented below has to be considered as one of a great many others. Our aim is to display the general character of the sound propagation. We start from (\ref{44}), insert the pressure values from (\ref{42}) and obtain $\sigma / 2 R_0=90$ MeV / fm$^3$. To remain within the interval $10\text{MeV/fm}^2 \leqslant \sigma \leqslant 120 \text{MeV/fm}^2[35]$ we choose $\sigma=90 \text{MeV/fm}^2$ and then $R_0=1$ fm which is somewhat too low according to \cite{35*}. With this value of $\sigma / R_0$ (\ref{45}) gives $\varphi=0.2$, $\Phi^2=0.8$. To get $\delta$ from (\ref{46}) we have to specify the value of $\eta$ from the extremely wide interval discussed above. Taking $\log _{10} \eta\,[\text{g/cm}\cdot\text{1}]=11$, or $\eta=18.7$ MeV/fm$^2$ one obtains $\delta=0.16$. The resonance frequency (\ref{47}) with the above values of parameters is $f_r=0.8 \cdot 10^{20}$ kHz ($\bar{\omega}_0=380$ MeV, $\varphi=0.2$).

\begin{figure}[h]
\begin{center}
\includegraphics[width=140mm,keepaspectratio=true]{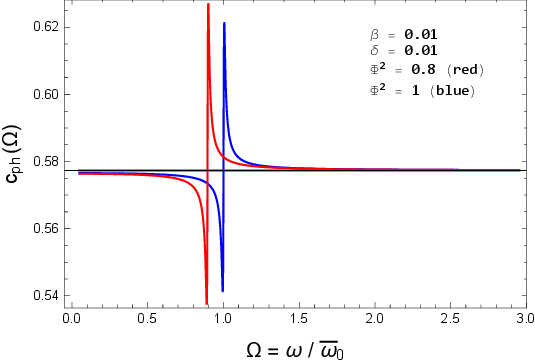}
\caption{The speed of sound in bubbly HNS as a function of $\Omega$ at $\beta=0.01$, $\delta = 0.01$. The red one corresponds for $\Phi^2=0.8$, and the blue one -- for $\Phi^2=1$. The green line is the sound speed with no bubbles.}
\end{center}
\label{figure02}

\end{figure}

The purpose of this Section was to demonstrate ties between different parameters and to show how to choose them in a self-consistent way.

\section{Results for the Sound Speed and Attenuation Coefficient}

Outside the region near the stability limit the sound speed and the attenuation coefficient are given by (\ref{33})-(\ref{34}). For the set of parameters (\ref{42})-(\ref{43}) $\varphi = 0.2 \ll 1$. 

In Fig.~1 we show the sound velocity defined by (\ref{33}) as a function of $\Omega=\omega / \bar{\omega}_0$ for the parameters (\ref{42})-(\ref{43}) and $\beta=0.1$, $\Phi^2=0.8\,(\varphi=0.2)$, $\delta=0.16$ (see the text). For comparison we present a similar curve with the same $\beta$ and $\delta$ but for $\Phi^2=1\,(\varphi=0)$. In Fig.~2 the same two curves are shown for $\beta=0.01$, $\delta=0.01$.  These Figs. (1 and 2) demonstrate the oscillatory behavior of the sound speed. This means that the sound propagation in a bubbly medium is highly nonlinear. According to (\ref{36}) at $\omega < \bar{\omega}_0 \Phi^2$ the sound velocity is reduced in comparison with the unperturbed HM value. This is because at low frequencies bubbles oscillate in phase with the driving sound wave and the compressibility increases. Above $\bar{\omega}_0 \Phi^2$ the bubble oscillations fall behind the oscillations of the driving wave and the medium becomes stiffer resulting in $c_{p h}>c_h$.

\begin{figure}[H]
\begin{center}
\includegraphics[width=140mm,keepaspectratio=true]{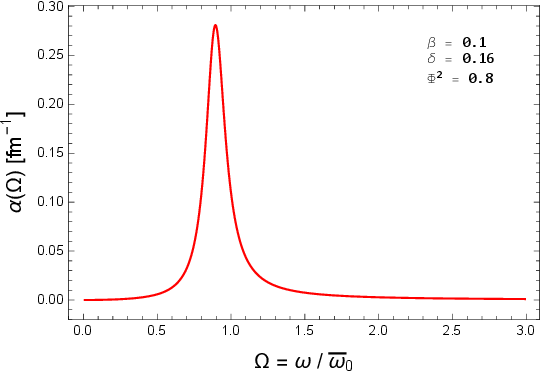}
\caption{The sound attenuation (\ref{34}) as a function of $\Omega$ at $\Phi^2=0.8$, $\beta=0.1$ and $\delta=0.16$.}
\end{center}
\label{figure03}
\end{figure}

Sound attenuation given by (\ref{34}) is presented in Fig.~3 as a function of $\Omega$ for $\Phi^2=0.8$, $\delta=0.16$ and (\ref{42})-(\ref{43}) for the values of other parameters. It has the $\omega^2$ dependence with an imposed resonance-like factor. Attenuation grows at or near the resonance frequency because the bubbles resonate thus causing the scattering and absorption of the sound wave. Generally, the $\omega^2$ dependence of $\alpha^{\prime}$ is not a universal law. The Mandelstam \!-\! Leontovich slow relaxation mechanism results either in linear $\omega$-dependence, or in frequency in dependence \cite{52,53,54,55}. Near the phase transition the sound absorption may be anomalously high \cite{60}.

\section{Summary and Discussion}

In this paper we presented a new view on the HNS with QM bubbles immersed into it. Such a star has very interesting acoustic properties. The presence of bubbles makes the sound speed highly dispersive and gives rise to additional attenuation. Even a small fraction of QM bubbles causes a high nonlinearity of the sound wave propagation. Our approach is based on the Rayleigh-Plesset hydrodynamical equation and on the polytropic EoS proposed in \cite{18}.

Equations were derived for the bubbly HNS compressibility (\ref{20}), the speed of sound (\ref{33}), the sound attenuation (\ref{34}) and the frequency of bubble pulsation at resonance (\ref{47}).

These equations include a set of NS parameters some and even most of them are hardly known. In particular this concerns the surface tension at the HM-QM interface and the HM shear viscosity. The bubble radius may be to some extent considered as a free parameter though according to a recent study the preferable value is $R_0 \sim \text{(5-10)\ fm}$ \cite{35*}. Being at first glance loosely determined the above parameters are firmly interrelated by Eqs. (\ref{42})-(\ref{43}) and (\ref{44})-(\ref{47}). Therefore the description of sound wave propagation in a bubbly HNS requires a self-consistent set of parameters. One of the possible patterns was presented in this work. Starting from the density and pressure values (\ref{42})- (\ref{43}) and imposing a restriction on the value of the surface tension we obtained $R_0 \simeq 1$ fm. According to (\ref{47}) the corresponding resonance frequency is $f_r \sim 10^{23}$ Hz. For $R_0 \simeq 100$ m (\ref{47}) yields $f_r \simeq 1$ MHz. All parameters are interrelated and one can not change $R_0$ keeping other quantities untouched. The search for a multitude of possible sets of parameters is beyond the scope of the present work.

A few words are needed to be added concerning the assumptions and limitations of our approach. These are:

\begin{enumerate}[label=\alph*)]
\item All bubbles are spherical and have the same equilibrium radius.
\item Bubbles occupy a small fraction of the total volume, $\beta \ll 1$.
\item Bulk (dilatational) viscosity effects of HM are negligible.
\item The sound wavelength $\lambda \gg R_0$.
\item The effects of bubble collapse, nucleation and percolation are beyond the scope of our study.
\end{enumerate}

The increase of $\beta$ would result in the interaction between the bubbles which is impartible to describe within the present approach. The requirement $\lambda \gg R_0$ allows to consider the driving sound wave as spatially homogeneous. The complete time-dependent solution of RPE beyond the linear approximation is possible only numerically \cite{a}. One may ask whether magnetic field of the order $B \simeq 10^{14}$ G in magnetars may substantially alter RPE. The answer is negative and magnetic field plays a minor role as compared with the shear viscosity \cite{46}.


Some interesting problems in HNSs acoustics remain for future studies. Among them is the sound propagation in QM-HM pasta with different geometrical configurations like rods, slaps, tubes. The problem of sound in layered media has been studied in detail in \cite{101}. High scattering cross section and sound wave reflection from the phase boundary were not discussed in the present study. The recent review of the nonlinear acoustics in matter with bubbles is presented in \cite{102}.


\section{Acknowledgments}

The authors are grateful to O.V. Rudenko for drawing our attention to interesting problems in bubbly acoustics. B.K. would like to thank Dmitry Voskresensky for useful remarks and discussion.


\end{document}